\documentclass[dvips]{acta}
\usepackage{supertabular,lscape,epsfig}
\usepackage{amssymb}
\usepackage{amsmath}
\SetPages{00}{000}
\SetVol{59}{2009}

\begin{document}
\renewcommand{\thefootnote}{\fnsymbol{footnote}}
\setcounter{footnote}{1}
\def\deg{$^{\rm o}$}
\begin{Titlepage}
\Title{A CCD Search for Variable Stars of Spectral Type B in the Northern\\ Hemisphere Open Clusters. VII. NGC\,1502}
\Author{G.~~M~i~c~h~a~l~s~k~a$^{1,2}$,~~~A.~~P~i~g~u~l~s~k~i$^2$,~~~M.~~S~t~\k{e}~\'{s}~l~i~c~k~i$^2$,~~~A.~~N~a~r~w~i~d$^2$}
{$^1$Universidad de Concepci\'on, Departamento de Astronom\'{\i}a,\\ Casilla 160-C, Concepci\'on, Chile\\
e-mail:gabi@astro-udec.cl\\
$^2$Instytut Astronomiczny Uniwersytetu Wroc{\l}awskiego,\\ Kopernika 11, 51-622 Wroc{\l}aw, Poland\\ 
e-mails:(pigulski,steslicki,narwid)@astro.uni.wroc.pl}
\Received{July 28, 2009}
\end{Titlepage}

\vspace*{-12pt}
\Abstract{We present results of variability search in the field of the young open cluster NGC\,1502. Eight
variable stars were discovered. Of six other stars in the observed field that were suspected for variability, we confirm variability of two, including 
one $\beta$~Cep star, NGC\,1502-26.  The remaining four suspects were found to be constant in our photometry. 
In addition, $UBVI_{\rm C}$ photometry of the well-known massive eclipsing binary SZ~Cam was obtained.  

The new variable stars include: two eclipsing binaries of which one is a relatively bright detached system with an EA-type light curve,
an $\alpha^2$\,CVn-type variable, an SPB candidate, a field RR Lyr star and three other variables showing variability of unknown origin. 
The variability of two of them is probably related to their emission in H$\alpha$, which has been measured by means of the $\alpha$ index 
obtained for 57 stars brighter than $V \approx$16 mag in the central part of the observed field.  Four other non-variable stars with emission
in H$\alpha$ were also found.

Additionally, we provide $VI_{\rm C}$ photometry for stars down to $V =$ 17~mag and $UB$ photometry for about 50 brightest stars in the 
observed field. We also show that the 10-Myr isochrone fits very well the observed color-magnitude diagram if a distance of 1~kpc and 
mean reddening, $E(V-I_{\rm C}) =$ 0.9~mag, are adopted.}
{stars: early-type -- stars: emission-line, Be --- open clusters and associations: individual: NGC\,1502}

\section{Introduction}
This is the seventh paper in the series in which we present results of the variability search for B-type variables 
(mainly $\beta$~Cep, SPB and massive binaries) in the young open clusters of the Northern Hemisphere. The results 
for previously studied clusters were briefly summarized by Pigulski \etal (2002).  In addition to the discovery or verification
of variability of about 70 stars, mostly members of the investigated clusters, it appeared that the incidence of 
$\beta$~Cep stars in the northern clusters is lower than in the southern ones. This was interpreted in terms 
of the metallicity gradient in the Galaxy. One of the main motivations for the search is also a selection of 
clusters containing large numbers of pulsators suitable for asteroseismology.  An example is the previous paper 
of the series (Ko\l{}aczkowski \etal 2004), in which we investigated 
the open cluster NGC\,6910 discovering four $\beta$~Cep stars. This prompted us to choose NGC\,6910 as a target of a two-year
photometric and spectroscopic campaign.  The data from the campaign are in the course of reduction, but preliminary analysis already 
increased the number of $\beta$~Cep stars in this cluster to seven (Pigulski \etal 2007) making it an excellent target for
seismic modelling and one of very few which are rich in $\beta$~Cep stars.

In the present paper, we report the results of the variability search in the field of another young open cluster, 
NGC\,1502. For the first time, the data from the new CCD camera, with a much larger field of view, are included.

\section{The Cluster}
NGC\,1502 (C\,0403$+$622, OCl\,383) is a loose young open cluster of Trumpler type II3p in Camelopardalis. Almost all studies
locate the cluster near the outer border of the Orion Spur at a distance of 0.7--1.0~kpc (Sanford 1949, Johnson \etal 1961, Purgathofer 1961,
Nicolet 1981, Reimann and Pfau 1987, Kharchenko \etal 2005, Paunzen \etal 2005), and only Tapia \etal (1991) derive a significantly larger distance
of 1.5~kpc. Because the main sequence of the cluster does not have a well-defined turnoff, the estimated age of NGC\,1502 has a large
spread, ranging from 5 to 15~Myr (Lindoff 1968, Harris 1976, Mermilliod 1981, Reimann and Pfau 1987, Tapia \etal 1991, 
Kharchenko \etal 2005).  Due to the similarity of ages and distances, a nearby ($\approx$6\deg apart) Cam OB1 association is frequently 
related to NGC\,1502 (de Zeeuw \etal 1999, Lyder 2001, Strai\v{z}ys and Laugalys 2007) and it is possible that both objects have 
a common origin. It was also suggested (Blaauw 1954, 1961, 1964; Stone 1991) that the O9.5\,Iae supergiant 
$\alpha$~Cam is a runaway star from NGC\,1502. The proper motions of stars in the field of the cluster were considered in many studies devoted to the
kinematics of the Galactic system of open clusters (see {\it e.g.}, Rastorguev \etal 1999), but some, like those of de Vegt (1966) or
Dias \etal (2001), may also be used to investigate the cluster membership.

The mean $E(B-V)$ color-excess of members of NGC\,1502 amounts to about 0.75~mag ({\it e.g.}, Reimann and Pfau 1987) but seems to vary 
in a range of the order of 0.2~mag across the cluster (Janes and Adler 1982, Yadav and Sagar 2001). However, as found by 
Tapia \etal (1991) and confirmed by Pandey \etal (2003) and Weitenbeck \etal (2008), the total-to-selective absorpion ratio
$R_V$ = $A_V/E(B-V)$ is for this region of the sky significantly smaller than the mean Galactic value 
of 3.1\,$\pm$\,0.1. The three studies provide $R_V$ between 2.4 and 2.7.  The study of Weitenbeck \etal (2008) was
based on polarimetric observations. They found that polarization for the cluster stars slightly exceeds 6\,\%, in accordance 
with earlier measurements (Hall and Mikesell 1950, Dombrovskii and Hagen-Thorn 1964).

The field of NGC\,1502 has been a subject of many photometric studies. The first two-color photographic photometry was published by Zug (1933).
The $UBV$ photometry, both photographic and photoelectric, was carried out by Hoag \etal (1961), Purgathofer (1964),
Dombrovskii and Hagen-Thorn (1964) and Reimann and Pfau (1987). Str\"omgren $uvby$ photometry was obtained by Reimann and Pfau 
(1987), Tapia \etal (1991), Delgado \etal (1992) and Crawford (1994). The three latter papers contain also $\beta$ 
photometry for some cluster members. The near-infrared $JHK$ photometry was provided by Tapia \etal (1991)
and observations in the narrow-band $\Delta a$ photometric system were published by Paunzen \etal (2005). 
The most thorough spectroscopic survey of the cluster was done by Huang and Gies (2006ab). They published MK spectral types, 
projected rotational velocities, radial velocities for two epochs, effective temperatures, surface gravities and helium abundances 
for 18 stars in the field of NGC\,1502. Some other authors published MK spectral types and/or radial velocities for 
several bright cluster members ({\it e.g.}, Hoag and Applequist 1965). The mean radial velocity of the cluster 
amounts to $-$21\,$\pm$\,6~km\,s$^{-1}$ (Kharchenko \etal 2005).

There are several numbering systems used for stars in the field of NGC\,1502 (Zug 1933, Hoag \etal 1961, Purgathofer 1964, Dombrovskii 
and Hagen-Thorn 1964, Tapia \etal 1991). Unless noted, we will follow the numbering provided by the WEBDA database.\footnote{The WEBDA 
database is available at {\tt http://www.univie.ac.at/webda/webda.html}.}

\begin{figure}[!ht]
\includegraphics[width=12cm]{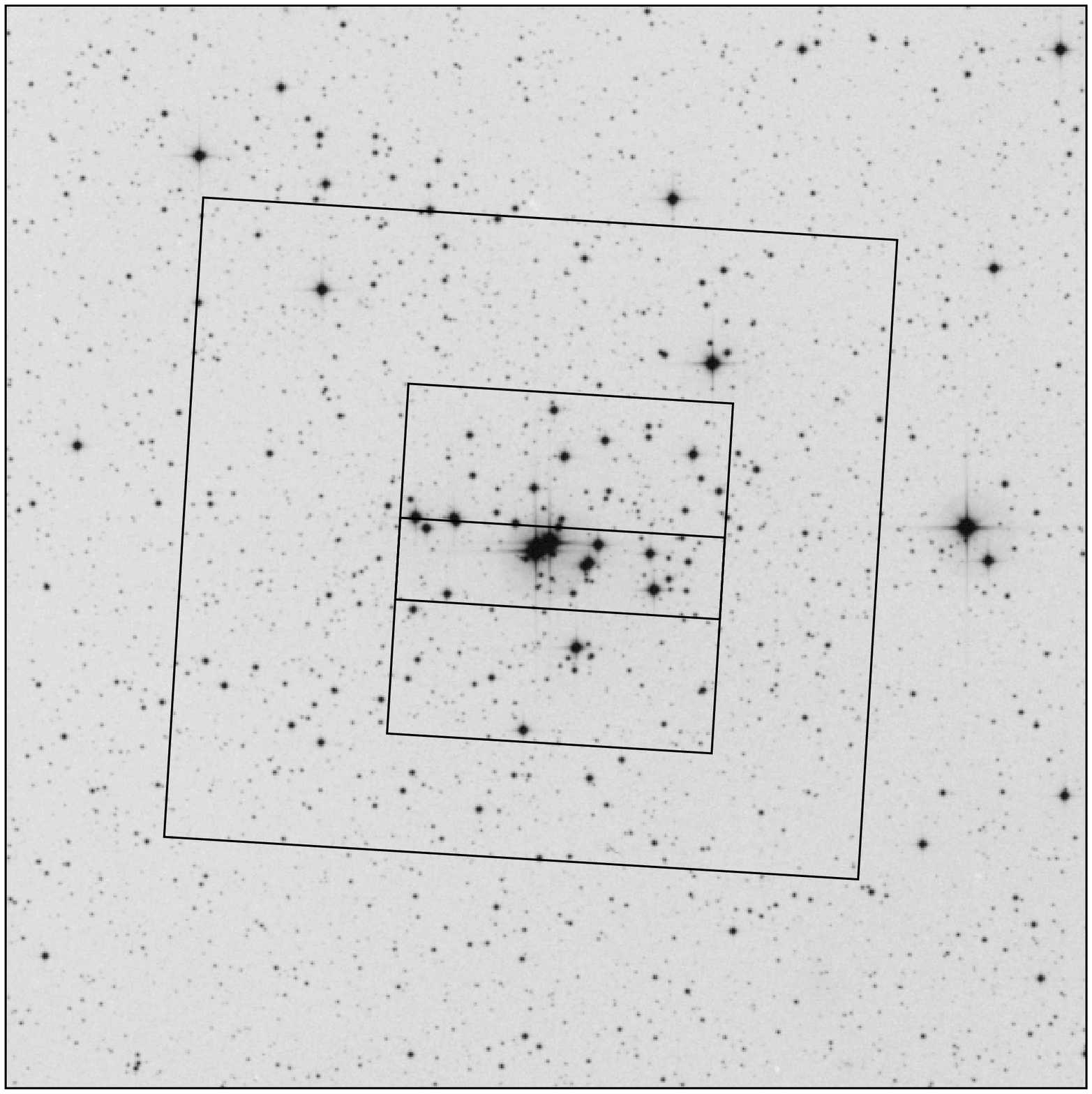}
\FigCap{A $20^{\prime}\times 20^{\prime}$ fragment of the POSS-II infrared plate centered on NGC\,1502. Solid 
lines delimit the fields observed with the two CCD cameras we used, see text for explanation.  North is
up, East, to the left.}
\label{poss}
\end{figure}

The most extensively studied star in NGC\,1502 is the eclipsing binary SZ Cam, the north-western component of the visual 
binary ADS\,2984, the bright pair in the center of the cluster (Fig.~\ref{poss}). SZ Cam = HD\,25638 = ADS\,2984B is itself 
a quadruple hierarchical system consisting of two spectroscopic binaries of which one is eclipsing.  The system will be presented in more detail 
in Section 4.3.  The other component of the pair, ADS\,2948A = HD\,25639, is also a spectroscopic binary.  
Several other stars in the field of NGC\,1502 were claimed to be variable or were suspected for variability. 
Guthnick and Prager (1930) noted photometric variability of HD\,25639 = NGC\,1502-1 with an amplitude of 0.05--0.06~mag. 
Next, two stars measured by Purgathofer (1964), NGC\,1502-44 (designated F in his paper) and 5, were labeled by him as `var?'. 
Hill (1967) observed photoelectrically 
six stars in NGC\,1502 and found variability with the $B$ amplitude of 18~mmag and a period of 0.190~d or 0.209~d 
in the combined light of one of them, namely 
a close visual pair, NGC\,1502-37 and 38. He concluded that one of the stars in the pair is a $\beta$~Cep-type candidate. 
Subsequently, from Str\"omgren photometry carried out on several nights, Delgado \etal (1992) found three variable candidates in NGC\,1502 
among 22 stars checked for variability. These were NGC\,1502-1, 26 and 61. NGC\,1502-26, showing frequencies of 10.5~d$^{-1}$ and 15.7~d$^{-1}$, 
appeared to be a very good $\beta$~Cep candidate. Another $\beta$~Cep candidate was
NGC\,1502-1. For this star, Delgado \etal (1992) found periodic variations with a frequency of 5.2~d$^{-1}$ superimposed on a variability on a longer time
scale, attributed to its binarity. The third star, NGC\,1502-61, they suspected to be variable on a long time scale.

The CCD photometry of NGC\,1502 we present here allows us to verify variability of all suspects and candidates mentioned above and carry out
a comprehensive variability study of stars in the cluster field. The results are discussed in Section 4; some preliminary findings were already 
published by St\k{e}\'{s}licki (2006).

\section{Observations and Reductions}
All observations of NGC\,1502 were carried out at Bia{\l}k\'ow, a station of University of Wroc{\l}aw, with a 60-cm 
Cassegrain telescope. During the first observing run, we acquired observations on 22 nights between August 22, 2004
and April 20, 2005. We used a 576\,$\times$\,384 pixels Star I CCD camera, the same which was used to obtain the data
for other clusters that were the subjects of previous papers of the series.  The camera covered a rectangle of only 6$^\prime \times$\,4$^\prime$.
For this reason, we observed two overlapping fields to obtain photometry of a larger number of cluster members. These fields are the two
overlapping rectangles shown in Fig.~\ref{poss}. The background image comes from the POSS-II\footnote{The Second Palomar Observatory 
Sky Survey (POSS-II) was 
made by the California Institute of Technology with funds from the National Science Foundation, the National Geographic Society, 
the Sloan Foundation, the Samuel Oschin Foundation, and the Eastman Kodak Corporation.} infrared plate centered on NGC\,1502.

Starting from the late summer of 2005, the observations of NGC\,1502 were resumed with the new, Andor DW432\,BV CCD camera having a 1250\,$\times$\,1152 
pixels back-illuminated chip. The use of the new camera resulted in a larger field of view, comprising about
13$^\prime \times$\,12$^\prime$ (large rectangle in Fig.~\ref{poss}). These observations were made on three nights, between September 8 
and October 7, 2005.  In addition, the cluster was observed 
on 12 nights between January 3 and March 30, 2008. In total, we gathered almost 14\,000 frames, which resulted in 162 hours
of time-series data (90 hours with old and 72 hours with the new camera). Observations were made primarily in Johnson $V$ and Cousins $I_{\rm C}$-bands. 
In 2008, some frames through Johnson $U$ and $B$ filters were also taken. Finally, like for most clusters we observed so far, H$\alpha$ 
images were secured for the purpose of finding stars with emission in this spectral line (see, e.g., Ko{\l}aczkowski \etal 2004). 

Depending on the camera and the sky conditions, the exposure times for time-series data were equal to 20--50 s for $V$ and 20--30 s for $I_{\rm C}$.
In order to increase dynamical range of the photometry, shorter exposures were also taken from time to time. This was done 
mainly in order to get better photometry for the two brightest stars in the field, the visual double ADS\,2984.
All frames were calibrated in a standard way and reduced by means of the {\sc DAOPHOT II} package (Stetson 1987).  Both profile 
and aperture photometries were obtained, the latter by means of the {\sc DAOGROW} program (Stetson 1990). 
We used both types of photometry: usually for bright unsaturated stars the aperture photometry appeared to have smaller scatter than
the profile photometry, while for the faint stars the opposite was true.

\section{Variable Stars}
As noted above, St\k{e}\'slicki (2006) already presented the results of the analysis of almost all frames of NGC\,1502 acquired with the old camera.
These results included the new data for SZ Cam = NGC\,1502-2, confirming variability of NGC\,1502-26, the $\beta$~Cep suspect, and
a discovery of two other variable stars, NGC\,1502-7 and 43. The former star was found to be eclipsing with an orbital period of 1.9839~d, 
while the variability of the latter was best represented by a sinusoid with a period of 0.41885~d.

The observations obtained with the new CCD camera not only cover much larger area of the cluster and its surroundings but also have a better quality.
For this reason, in the search for variability we used primarily the data from the new camera. However, a search for variable stars
was also carried out using the data obtained with the old camera in order to be sure that no transient changes or eclipses were missed. 
In the final analysis, data from both cameras were combined for stars found to be variable.
\begin{figure}[htb]
\includegraphics[width=12cm]{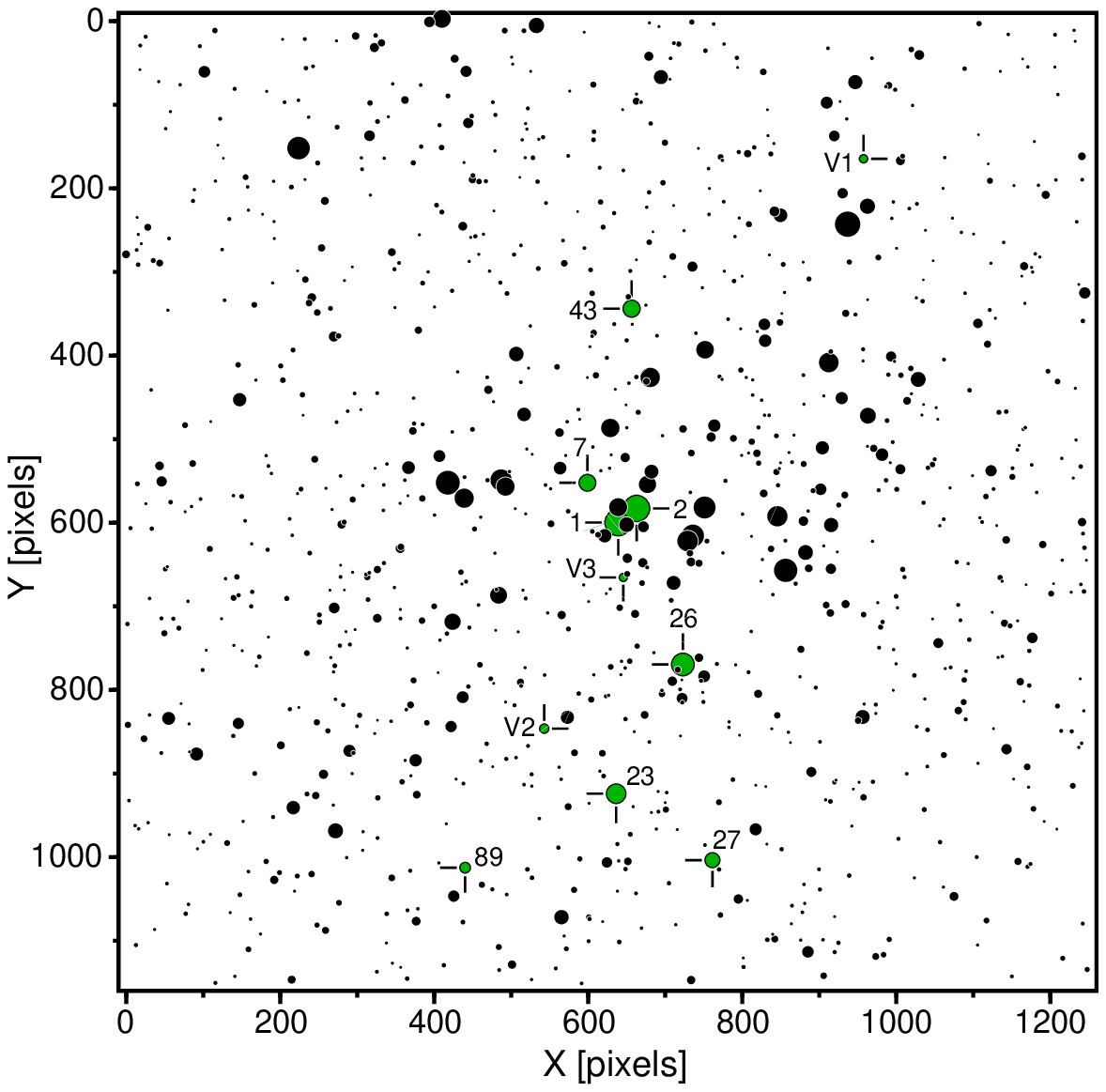}
\FigCap{Schematic map of the observed field with positions of all stars detected in the $I_{\rm C}$ band. Variable stars are marked and
labeled (see Table 1 for designations). North is approximately up, East, to the left.}
\label{map}
\end{figure}

Comparison stars were chosen in a trial-and-error procedure out of the bright unsaturated stars. The differential magnitudes 
were analyzed by means of a Fourier periodogram routine in the range between 0 and 40 d$^{-1}$. Then, the periodograms, 
the light curves and phase diagrams were inspected by eye and the final decision as to the variability was taken.  Among 1034 stars 
detected in the field of view of the new camera eleven showed variability.  They are indicated in the schematic map 
in Fig.~\ref{map} and listed in Table 1. Three faintest variables that do not have numbers in the WEBDA database 
were designated as V1, V2, and V3. In addition, we list in Table 1 the four other suspects (NGC\,1502-61, 37, 44 and 5)
that appeared to be constant in our photometry. For these stars, the detection threshold in mmag is given in the fourth column of Table 2 
instead of the variability ranges.

\MakeTable{crcclcl}{10cm}{The variable stars and the suspected variable stars in NGC\,1502}
{\hline\noalign{\vskip1pt}
WEBDA&\multicolumn{1}{c}{$V$}&\multicolumn{1}{c}{$V-I_{\rm C}$}&$\Delta V$, $\Delta I_{\rm C}$& 
\multicolumn{1}{c}{Period}&Cluster&Type of variability,\\
number& \multicolumn{1}{c}{[mag]}&\multicolumn{1}{c}{[mag]}&[mag]&\multicolumn{1}{c}{[d]}&member&notes\\
\noalign{\vskip1pt}
\hline
\noalign{\vskip1pt}
1 &  6.91 & 0.65 & 0.08,0.07 & --- &yes  &unknown (NSV\,1458)\\ 
2 &  6.91 & 0.61 & 0.30,0.28 & 2.698415(7) & yes &EA (SZ Cam)\\ 
26&  9.63 & 0.74 & 0.007,0.006 & 0.09611475(8) & yes &$\beta$~Cep (NSV\,15899)\\  
23& 10.69 & 0.81 & 0.08,0.09 & --- & yes &Be \\  
43& 11.36 & 0.72 & 0.016,0.014 & 0.68613(9) & yes &Ell or SPB\\ 
7 & 11.49 & 0.76 & 0.16,0.16 & 1.984476(6)  & yes &EA \\ 
27& 12.42 & 0.81 & 0.07,0.08 & 3.3085(4) & yes & $\alpha^2$\,CVn \\  
89& 14.65 & 1.47 & 0.05,0.03 & 0.74266(11) & uncert. & unknown\\ 
V1& 15.45 & 1.31 & 0.52,0.44 & 0.2992291(4)& no &RRc\\   
V2& 15.47 & 1.61 & 0.11,0.06 & 1.00695(12) & uncert.&$\lambda$~Eri?\\ 
V3& 15.91 & 1.42 &  --- ,0.28 & 1.76845(13) & uncert.&EW?\\ 
\noalign{\vskip1pt}\hline\noalign{\vskip1pt}
61&  9.56 & 0.63 & $<$2.7,$<$3.3 & --- & yes & cst (NSV 15901)\\ 
37&  9.65 & 0.70 & $<$1.5,$<$1.6 & --- & yes & cst (NSV 15898)\\ 
44& 10.76 & 0.75 & $<$2.4,$<$2.6 & --- & yes & cst (NSV 1456)\\ 
5 & 11.41 & 0.72 & $<$6,$<$6 & --- & yes & cst (NSV 1457)\\ 
\noalign{\vskip1pt}
\hline
\label{vtab}
}

\subsection{$\beta$~Cep star NGC\,1502-26 (NSV\,15899)}
\begin{figure}[!ht]
\includegraphics[width=11.5cm]{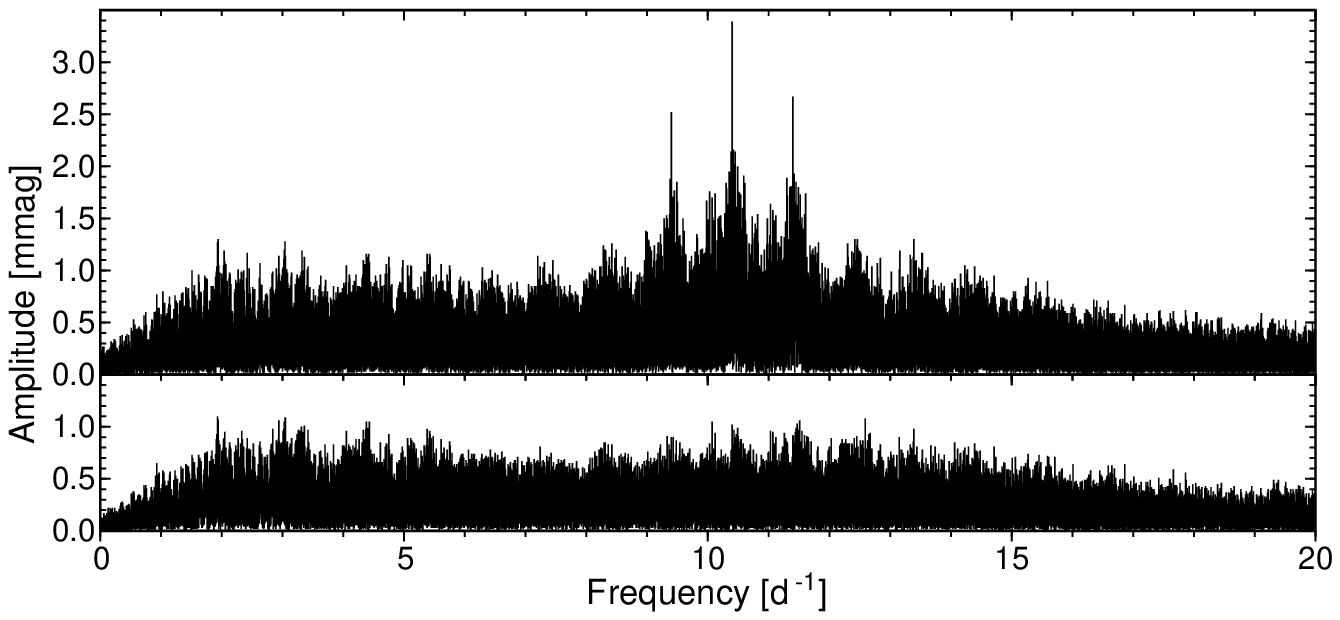}
\FigCap{{\it Top}: Fourier amplitude spectrum of the combined 2005--2008 $V$-filter data of NGC\,1502-26. {\it Bottom}: The same after prewhitening
with the mode having frequency of 10.404230~d$^{-1}$.}
\label{var6}
\end{figure}

As noted above, NGC\,1502-26 was the best $\beta$~Cep candidate in NGC\,1502, discovered by Delgado \etal (1992) and later confirmed by
St\k{e}\'slicki (2006) using data obtained with the old camera. The observations carried out with the new camera lowered the detection threshold,
but the mode with frequency of 10.404~d$^{-1}$ remained the only one detected in the photometry of this star.  It was found in the $B$, $V$, 
and $I_{\rm C}$-filter data, but not in $U$; apparently, the accuracy of the $U$ photometry was insufficient to allow detection.
The Fourier amplitude spectrum for NGC\,1502-26 $V$-filter combined 2005--2008 data is shown in Fig.~\ref{var6}. Decreased noise
in the low-frequency domain is due to detrending. It was done in the following way.  After removing the periodic signal, the nightly means 
were calculated using residuals.  These means were subsequently interpolated by a cubic spline and such smoothed curve was subtracted 
from the original data.

The two radial-velocity observations obtained for this star by Huang and Gies (2006a) differ by about 35 km/s. 
They concluded therefore that the star 
is a single-lined spectroscopic binary. This is still a possibility, but a pulsation is a much more likely cause of the difference.  
The projected rotational velocity derived by the same authors amounts to 174\,$\pm$\,8~km/s; 
the high rotational velocity of NGC\,1502-26 is consistent with the MK spectral type, B1.5\,Vn, reported by Hoag and Applequist (1965).

The parameters of the sine-curve fit for the single mode with a frequency of 10.404230\,$\pm$\,0.000009~d$^{-1}$, 
found in NGC\,1502-26, are given in Table 2. The columns are self-explanatory;
we only note that observations with the old and new camera are designated as `O' and `N', respectively. Numbers in the
parentheses are the standard deviations with the leading zeroes suppressed. The other 
periodicity suggested by Delgado \etal (1992) for NGC\,1502-26, having frequency of 15.7~d$^{-1}$, was not found in our data.

\MakeTable{crcrrr}{10cm}{Parameters of the sine-curve fit for NGC\,1502-26. $T_0$ = HJD\,2\,453\,000}
{\hline\noalign{\vskip1pt}
Filter &$N_{\rm obs}$ & \multicolumn{1}{c}{Semi-amplitude} & \multicolumn{1}{c}{$T_{\rm max} - T_0$} & $S/N$ &\multicolumn{1}{c}{$\sigma_{\rm res}$} \\
(camera)& & \multicolumn{1}{c}{[mmag]} & \multicolumn{1}{c}{[d]}&&\multicolumn{1}{c}{[mmag]}  \\
\noalign{\vskip1pt}
\hline
\noalign{\vskip1pt}
$B$ (N) & 762 &  3.39(28) & 1487.2497(12) & 6.2&5.40\\
\noalign{\vskip1pt}
$V$ (O) &2402 &   3.38(15) &  394.9056(07) &12.6&5.01\\
$V$ (N) &1498 &   3.80(19) & 1360.6650(08) & 8.1&5.15\\
$V$(O+N)&3900 &   3.63(12) &  765.8114(05) &14.5&5.07\\
\noalign{\vskip1pt}
$I_{\rm C}$ (O) & 652 &  3.30(38) &  361.6477(20) & 5.6&7.16\\
$I_{\rm C}$ (N) &1693 &  3.09(15) & 1489.8437(08) & 8.5&4.46\\
$I_{\rm C}$ (O+N)&2345 &  3.12(16) & 1176.1251(08) &10.0&5.35\\
\noalign{\vskip1pt}
\hline
\label{sfit}
}

\begin{figure}[!ht]
\includegraphics[width=12cm]{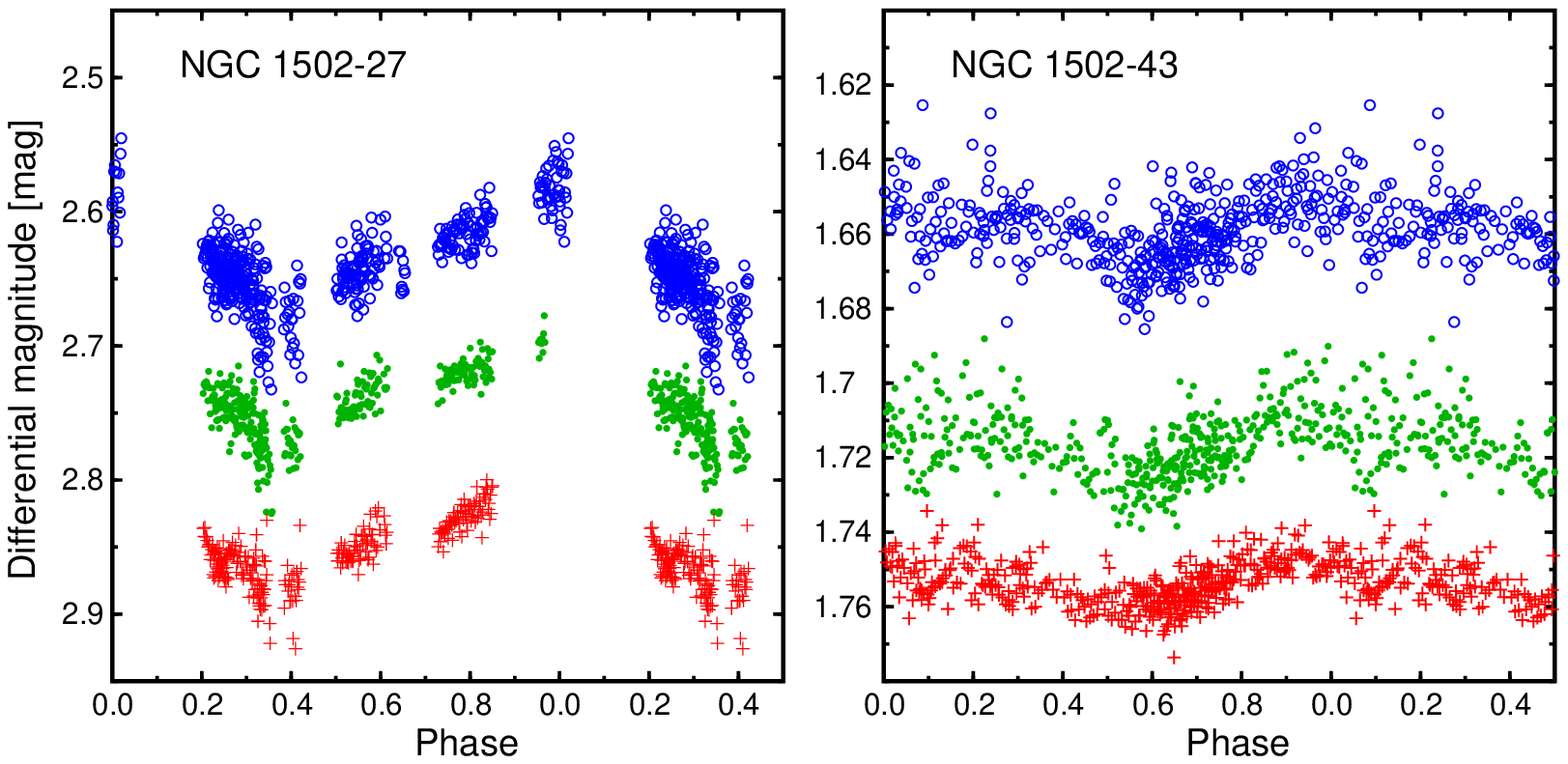}
\FigCap{{\it Left:} The averaged differential $B$ (open circles), $V$ (dots) and $I_{\rm C}$ (plus signs) 
observations of the $\alpha^2$\,CVn-type star NGC\,1502-27 phased with the period of 3.3085~d. {\it Right:} The same for the 
SPB suspect, NGC\,1502-43, phased with a period of 0.68613~d. In both panels offsets were applied in order to separate 
light curves in different bands.}
\label{var27}
\end{figure}

\subsection{Two Periodic B-Type Variables}
Two other early-type members of the cluster appear to be periodic variables. NGC\,1502-27 was observed only with the new camera. 
The variability of this star is slightly non-sinusoidal in shape (Fig.~\ref{var27}). 
The three 2005 nights, on which only $V$ observations were made, indicate that the amplitude was smaller in 2005 than in 2008. 
The location of NGC\,1502-27 in the color-color diagram (Fig.~\ref{ccd}) is consistent with a cluster late B-type star, that matches 
the Harvard classification of B8 given by Zug (1933). The star was found to be an Ap star by means of the Geneva photometry by
North and Cramer (1981). Its peculiarity was recently confirmed by Paunzen \etal (2005) who used $\Delta a$ photometry. This,
as well as the period of variability ($\approx$3.3 d), show that the star is an $\alpha^2$CVn-type variable. 
The star appears in the catalogue of Ap and Am stars as Renson 6525 (Renson and Manfroid 2009).

The only SPB candidate we found is NGC\,1502-43, a B6-type star according to Dombrovskii and Hagen-Thorn (1964). The light variability of this
star can be represented by a single sinusoidal term with a period of about 0.687~d (Fig.~\ref{var27}) and semi-amplitudes equal to 
7.0\,$\pm$\,0.5, 7.5\,$\pm$\,0.6 and 5.9\,$\pm$\,0.4~mmag in $B$, $V$, and $I_{\rm C}$, respectively.  Huang and Gies (2006a) report two
measurements of radial velocity for this star; they differ by about 100~km/s.  In view of the small photometric amplitudes, it seems rather
unlikely that the radial velocity difference is due to pulsations.  Although pulsation as the cause
of variability cannot be rejected, the more plausible explanation of the observed light and 
radial velocity variations is that NGC\,1502-43 is a non-eclipsing spectroscopic binary showing low-amplitude photometric modulations due to
proximity effects.  In other words, it is rather an ellipsoidal variable than an SPB star. If this were indeed the case, the period given in Table 1 
would be equal to half the orbital period.

\begin{figure}[!ht]
\includegraphics[width=12.5cm]{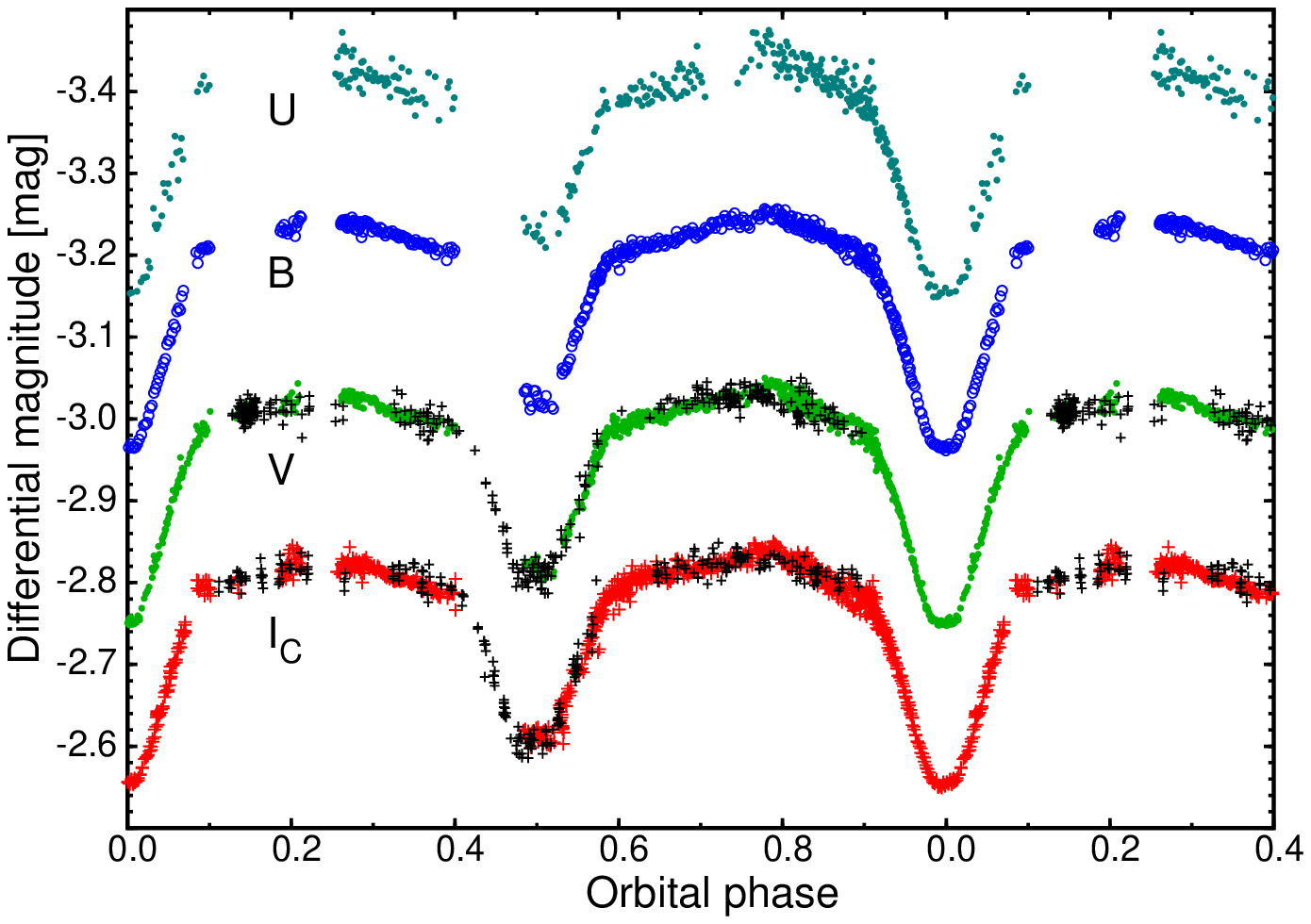}
\FigCap{Phase diagram of the $U$, $B$, $V$ and $I_{\rm C}$ observations of the eclipsing system SZ\,Cam (NGC\,1502-2). The $V$ and $I_{\rm C}$ observations 
from the old CCD camera are shown with plus signs. Phase 0.0 corresponds to the primary minimum. For clarity, shifts between light curves 
in different bands were applied.}
\label{sz-cam}
\end{figure}

\subsection{SZ Camelopardalis}
SZ Cam (NGC\,1502-2, HD\,25638, BD$+$61{\deg}676\,N) is the brightest member of the cluster and the B component of
the visual double ADS\,2984 (separation 18$^{\prime\prime}$). The star was discovered as a spectroscopic binary by Plaskett (1924).
Subsequently was found to be an eclipsing binary (Guthnick and Prager 1930), but the first comprehensive photometric 
study of the system was published by Wesselink (1941). This author was the first to derive the true orbital period, equal to 2.698~d. 
These early works were followed by many other studies (Kopal and Shapley 1956, Heintze and Grygar 1970, 
Kitamura and Yamasaki 1972, Budding 1974, 1975, 1977, Mardirossian \etal 1980) in which the authors claimed that the system has a semi-detached 
configuration with a mass ratio of 0.2--0.3.
Consequently, the increase of the orbital period found by Chochol (1977, 1980) was interpreted in terms of mass exchange between
components. However, observations made with a higher spectral resolution (Mayer \etal 1994, Lorenz \etal 1998, Harries \etal 1998) allowed 
a better separation of the lines of the components resulting in improved, much higher mass ratio, close to 0.7. The new solutions showed the system
to be detached. Earlier, Mayer (1987) interpreted the changes of the orbital period of SZ\,Cam in terms of the light-time effect
thus suggesting also the existence of a tertiary component in the system. The spectral lines of the tertiary were 
indeed found in the spectra obtained by Mayer \etal (1994)
and the tertiary appeared to be itself a single-lined spectroscopic binary having an orbital period of 2.797~d (Mayer \etal 1994, Gorda 2002,
Michalska \etal 2007). 
The wide pair orbits the center of mass of the quadreple system in an eccentric ($e =$ 0.78) orbit with a period of about 50~yr 
(Mayer \etal 1994, Lorenz \etal 1998, 
Gorda \etal 2007, Gorda 2008). The pair was also resolved by means of speckle interferometry (Mason \etal 1998, Gorda \etal 2007). A combination of
the visual and spectroscopic (from the study of the light-time effect) orbits allowed to derive the distance to SZ~Cam, 1125\,$\pm$\,135~pc (Gorda \etal
2007), in a reasonable agreement with the determinations of cluster distance based on photometric methods and spectroscopic parallax. 
The total mass of the NGC\,1502-2 quadruple system exceeds 50~M$_\odot$. The eclipsing pair is classified as O9\,IV + B0.5\,V (Lorenz \etal 1998). 

The $UBVI_{\rm C}$ light curves of SZ~Cam from our observations are shown in Fig.~\ref{sz-cam}. We have derived 14 epochs of minimum light which
are listed in Table 3.  They were obtained by a least-squares fit of a smoothed light-curve to data subsets covering an epoch of minima.
The method uses the whole light curve, thus allowing determination of an epoch of minimum light even if it was only partially covered by observations.
The derived epochs of minimum light were used to obtain the following ephemeris:
$$\mbox{Min I = HJD\,2453355.0682(21) + 2.698415(7)} \times E,$$
where $E$ is the number of the orbital cycles elapsed from the initial epoch and the numbers in parentheses are the {\it rms} errors with leading
zeroes omitted. We would like to point out, however, that the orbital period of SZ\,Cam 
varies due to the light-time effect, so that over a longer time scale the ephemeris given by Gorda \etal (2007) should be used.

\MakeTable{rcrcrcr}{10cm}{Epochs of minimum light for SZ Cam and NGC\,1502-7}
{\hline\noalign{\vskip1pt}
\multicolumn{3}{c}{SZ Cam}&&\multicolumn{3}{c}{NGC\,1502-7}\\
\noalign{\vskip1pt}
Filter & Minimum&\multicolumn{1}{c}{$T_{\rm max} -$\mbox{HJD\,2453000.0}}&\qquad& Filter &Minimum& \multicolumn{1}{c}{$T_{\rm max}-$\mbox{HJD\,2453000.0}} \\
\noalign{\vskip1pt}\hline\noalign{\vskip1pt}
$V$&sec.&356.4167(16)&&  $V$&prim.&350.7594(05)\\
$I_{\rm C}$&sec.&356.4090(41)&&  $V$&prim.&408.3038(05)\\
$V$&sec.&410.3832(09)&&  $I_{\rm C}$&prim.&408.3025(14)\\
$I_{\rm C}$&sec.&410.3785(14)&&  $V$&prim.&410.2904(05)\\
$V$&sec.&429.2812(17)&&  $I_{\rm C}$&prim.&410.2907(11)\\
$I_{\rm C}$&sec.&429.2747(64)&&  $V$&prim.&622.6299(07)\\
$V$&prim.&649.2037(11)&&  $B$&prim.&1521.6051(15)\\
$U$&prim.&1469.5114(33)&&  $V$&prim.&1521.6094(16)\\
$B$&prim.&1469.5135(08)&&  $I_{\rm C}$&prim.&1521.6089(07)\\
$V$&prim.&1469.5115(09)&&  $B$&prim.&1553.3365(35)\\
$I_{\rm C}$&prim.&1469.5118(12)&&  $V$&prim.&1553.3365(19)\\
$U$&sec.&1511.3323(45)&&  $I_{\rm C}$&prim.&1553.3462(10)\\
$B$&sec.&1511.3457(20)&&  $B$&prim.&1555.3215(14)\\
$I_{\rm C}$&sec.&1511.3423(20)&&  $V$&prim.&1555.3219(15)\\
&&&&  $I_{\rm C}$&prim.&1555.3249(09)\\
\noalign{\vskip1pt}
\hline
\label{mint}
}

\begin{figure}[!ht]
\includegraphics[width=12cm]{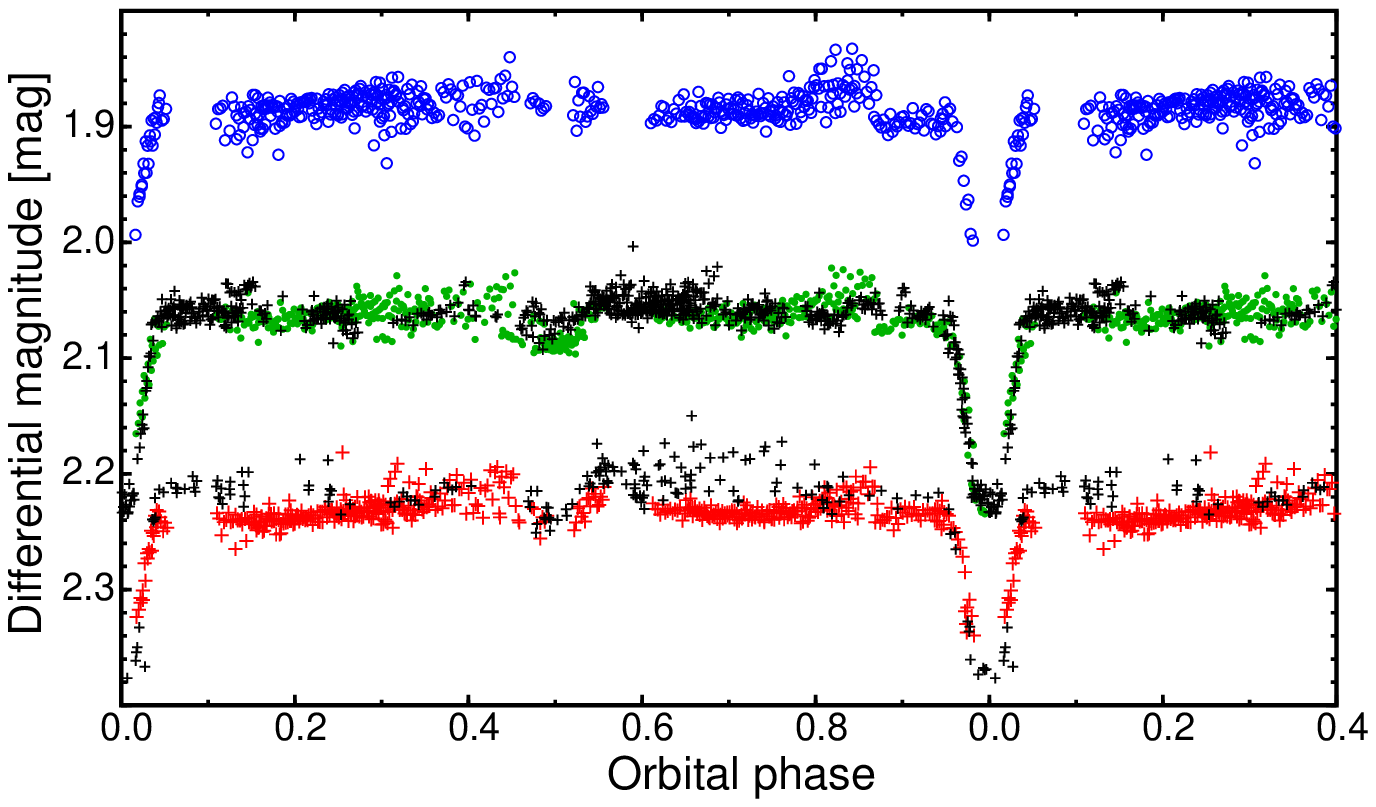}
\FigCap{The $BVI_{\rm C}$ (from top to bottom) observations of NGC\,1502-7 phased with the period of 1.984476~d. Some offsets were applied
to separate light curves in different bands. Phase 0.0 corresponds to the primary minimum.}
\label{var7}
\end{figure}

\subsection{Other Eclipsing Binaries}
NGC\,1502-7 appear to be another EA-type eclipsing binary. It is also a likely member of the cluster. The times of the primary minimum
are listed in Table 3. They were derived in the same way as for SZ~Cam and resulted in the following ephemeris:
$$\mbox{Min I = HJD\,2453350.7565(20) + 1.984476(6)} \times E.$$
The light curves of NGC\,1502-7 (Fig.~\ref{var7}) show a
0.18 mag deep primary minimum with a flat bottom and a very shallow ($\approx$0.02 mag) secondary minimum.
This shows that the secondary component is much fainter and smaller than the primary. The system is detached; the components
are probably main-sequence stars. Unfortunately, the two radial-velocity measurements for NGC\,1502-7 given by Huang and Gies (2006a)
are separated by almost exactly one orbital period; this does not allow to estimate the range of radial velocities for the system's primary.

\begin{figure}[!ht]
\includegraphics[width=10cm]{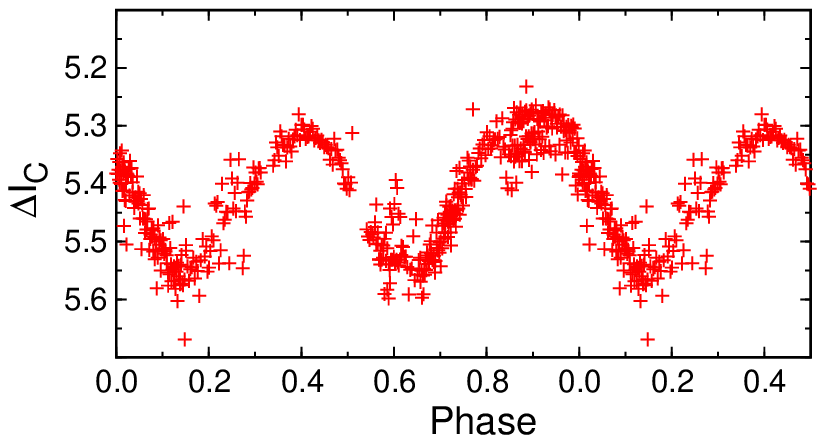}
\FigCap{Phase diagram of the 2008 $I_C$ observations of V3, possible eclipsing EW-type system.  Phase 0.0 was chosen arbitrarily.}
\label{star121}
\end{figure}

Another variable star that might be an eclipsing binary is V3, a relatively faint star located close to the cluster center.  
The proximity of the brightest stars in the field  
affected the photometry of V3 so much that the variability can be clearly seen only in the 2008 $I_{\rm C}$ data and 
barely in the 2008 $V$-filter photometry.  The shape of the light curve (Fig.~\ref{star121}) 
resembles W\,UMa-type variability. There is also a weak evidence for the O'Connell effect. However, the period (1.77~d) is rather long for this
type of eclipsing binary. Therefore, we cannot exclude the possibility that the true period is half that given in Table 1 and that the
cause of variability is other than eclipses.

\begin{figure}[!ht]
\includegraphics[width=12cm]{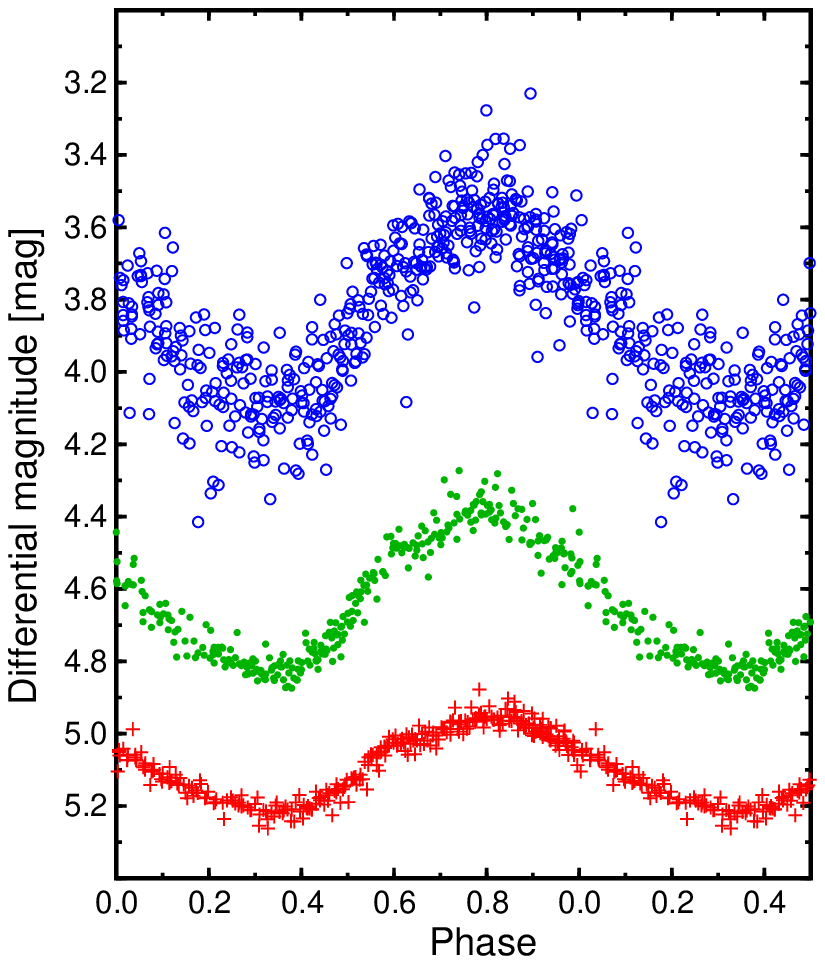}
\FigCap{The averaged differential $B$ (open circles), $V$ (dots) and $I_{\rm C}$ (plus signs) observations of the background RR Lyr 
star V1 phased with the period of 0.2992291~d. Offsets were applied to separate light curves in different bands.}
\label{rrl-116}
\end{figure}

\subsection{The RR Lyrae Star V1}
V1 is clearly an RRc-type star judging from the period, shape of the light curve and increasing amplitude when 
going from $I_{\rm C}$ to $B$ (Fig.~\ref{rrl-116}).  The star is definitely not a member because the cluster is much too young to contain
such an evolved star. Adopting the instrinsic $(V-I_{\rm C})_0 =$ 0.35~mag for an RRc star (see {\it e.g.}, Kopacki 2001), we get $E(V-I_{\rm C}) \approx$ 
0.99~mag corresponding to $E(B-V) \approx$ 0.8~mag if color excess ratio given by Winkler (1997) is applied. 
Adopting $R_V =$ 2.6 (see Section 2) we get
$A_V \approx$ 2.1~mag. The absolute magnitude of the star cannot be estimated accurately because its metallicity is not known. However,
assuming a typical value of [Fe/H] $=-$1.5 and using the relation given by Olech \etal (2003), we get an estimate of the true distance
modulus, 12.8~mag, and a distance of 3.5~kpc. V1 is therefore a background object, only slightly more reddened than the cluster, 
but located at a much larger distance than NGC\,1502.

\begin{figure}[!ht]
\includegraphics[width=12cm]{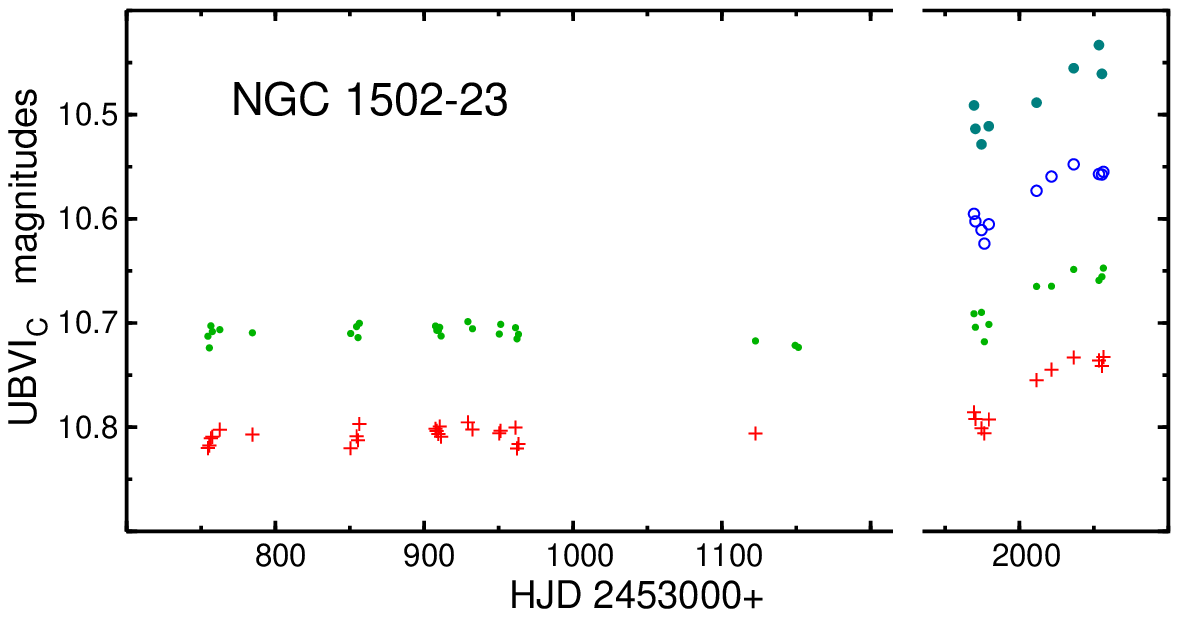}
\FigCap{$UBVI_{\rm C}$ (from \textit{top} to \textit{bottom}) light curves of NGC\,1502-23. The following offsets in magnitudes have been applied: $U$: $-$0.78~mag,
$B$: $-$0.75~mag, $V$: no offset, $I_{\rm C}$: $+$0.9~mag. Each point corresponds to a nightly mean.}
\label{var23}
\end{figure}

\subsection{Variable Stars Showing H$\alpha$ Emission}
Fast rotation and the presence of circumstellar disks in Be stars result in variability on different time scales, both regular
and irregular. Of the stars with emission we detected by measuring the $\alpha$ index (Section 6), two appeared to be variable. 
NGC\,1502-23 showed irregular behavior (Fig.~\ref{var23}) typical of Be stars. Its MK spectral type as determined by Hoag and Applequist
(1965) is B3\,V; the
location in the two-color diagram (Fig.~\ref{ccd}) is consistent with this classification.  

\begin{figure}[!ht]
\includegraphics[width=12cm]{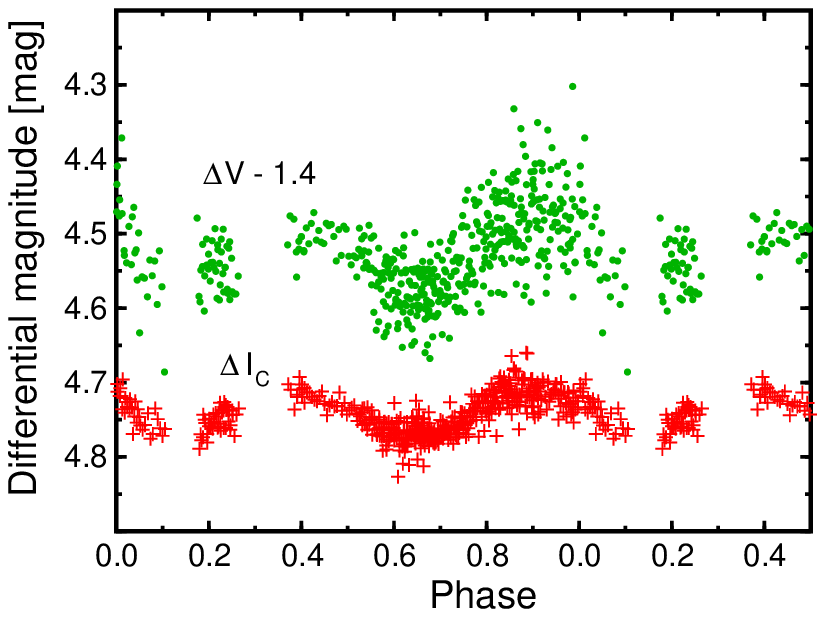}
\FigCap{$VI_{\rm C}$ light curves of V2 phased with the period of 1.00695~d.}
\label{var-o98}
\end{figure}
Another star with H$\alpha$ emission, V2,
is also a variable. The dominant periodicity has a period of about 0.503~d or, more likely, twice this value, 1.007~d (Fig.~\ref{var-o98}). 
The proximity of the period to 1~d makes difficult the derivation of an unambiguous period from the combined 2005 and 2008 data.
In addition to the period given in Table 1, another variation was found in the $I_{\rm C}$-filter data, with a period of 
1.966~d, \textit{i.e.}, about twice as long as the dominant period. Fig.~\ref{var-o98} shows the $VI_{\rm C}$ data for this star, 
averaged over 0.0025-day intervals and phased with the 1.007-day period. As the spectral type of this star is not known, 
it is difficult to assign the type of variability, but the period and the H$\alpha$ emission indicate a $\lambda$~Eri type.

\subsection{Other Variables}
\begin{figure}[!ht]
\includegraphics[width=12cm]{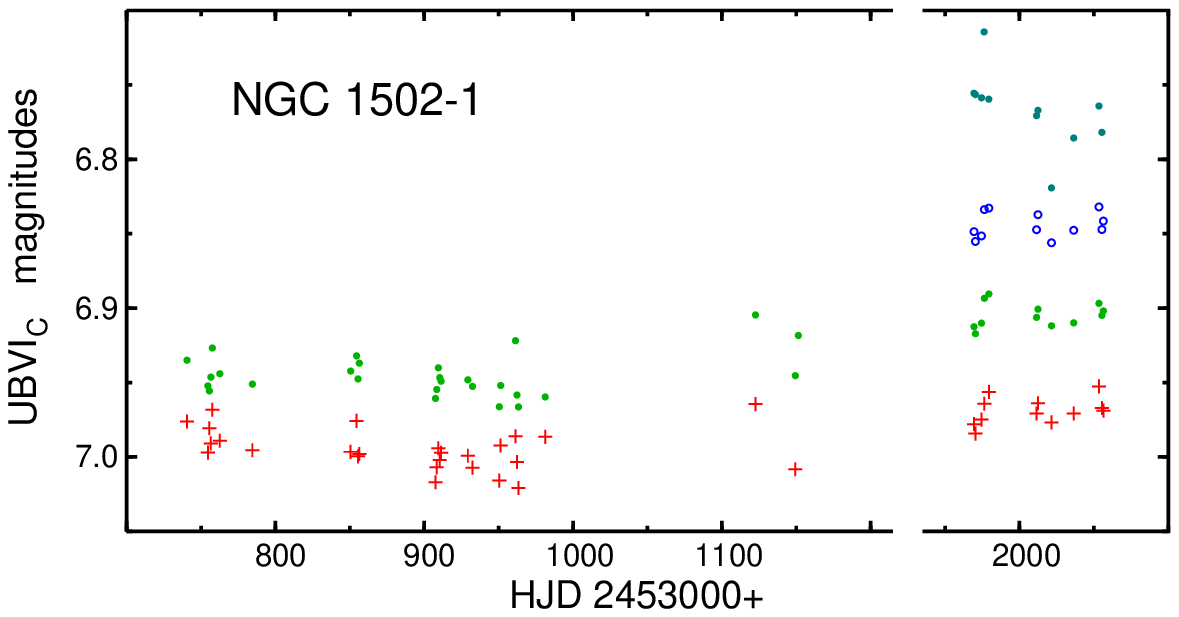}
\FigCap{$UBVI_{\rm C}$ (from \textit{top} to \textit{bottom}) light curves of NGC\,1502-1. The following offsets in magnitudes were applied: $U$: $-$0.25~mag,
$B$: $-$0.55~mag, $V$: no offset, $I_{\rm C}$: $+$0.72~mag.}
\label{var1}
\end{figure}

As noted in Section 2, NGC\,1502-1 = ADS\,2948A = NSV\,1458 was claimed to be variable by Guthnick and Prager (1930). The star has a spectral type
of B0\,II-III (Morgan \etal 1955) and is known to have variable radial velocity (Plaskett 1924, Lorenz \etal 1998, Hohle \etal 2009). 
Unfortunately, the photometric 
variability of NGC\,1502-1 (Fig.~\ref{var1}) is rather erratic.  We also analyzed archival radial velocities in the hope of 
finding periodic variations. Unfortunately,
due to the scarcity of the radial velocity data, the results were inconclusive.

\begin{figure}[!ht]
\includegraphics[width=12cm]{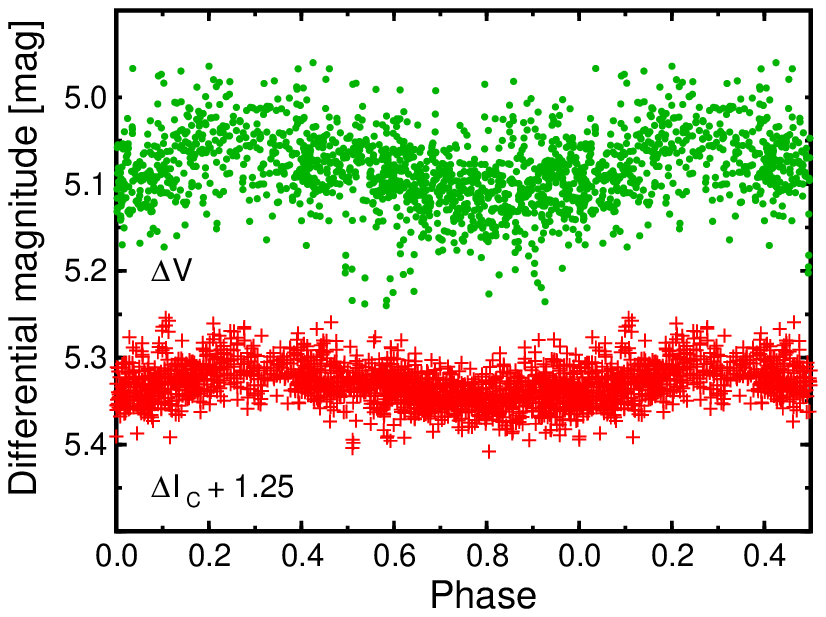}
\FigCap{The differential 2008 $V$ (dots) and $I_{\rm C}$ (plus signs) observations of NGC\,1502-89
phased with the period of 0.74266~d.}
\label{var89}
\end{figure}

Finally, periodic low-amplitude variations were found in the photometry of NGC\,1502-89 (Fig.~\ref{var89}). 
The period is equal to 0.74266~d.  However, the spectral type of this star is not known and consequently no final type of variability can be assigned.

As mentioned in Section 2, four other stars in the field, NGC\,1502-61, 37, 44 and 5, were claimed to be photometrically variable. 
We do not find any evidence for their 
variability with semi-amplitudes higher than the values given in the fourth column of Table 1. Unlike for the variable stars,
the detection thresholds for these stars, corresponding to $S/N =$ 4 in the Fourier periodogram, are given in mmag.

The time-series photometry of all variable stars we found in NGC\,1502 has
been deposited in the {\it Acta Astronomica Archive}.

\section{$UBVI_{\rm C}$ Photometry}
\begin{figure}[!ht]
\includegraphics[width=11cm]{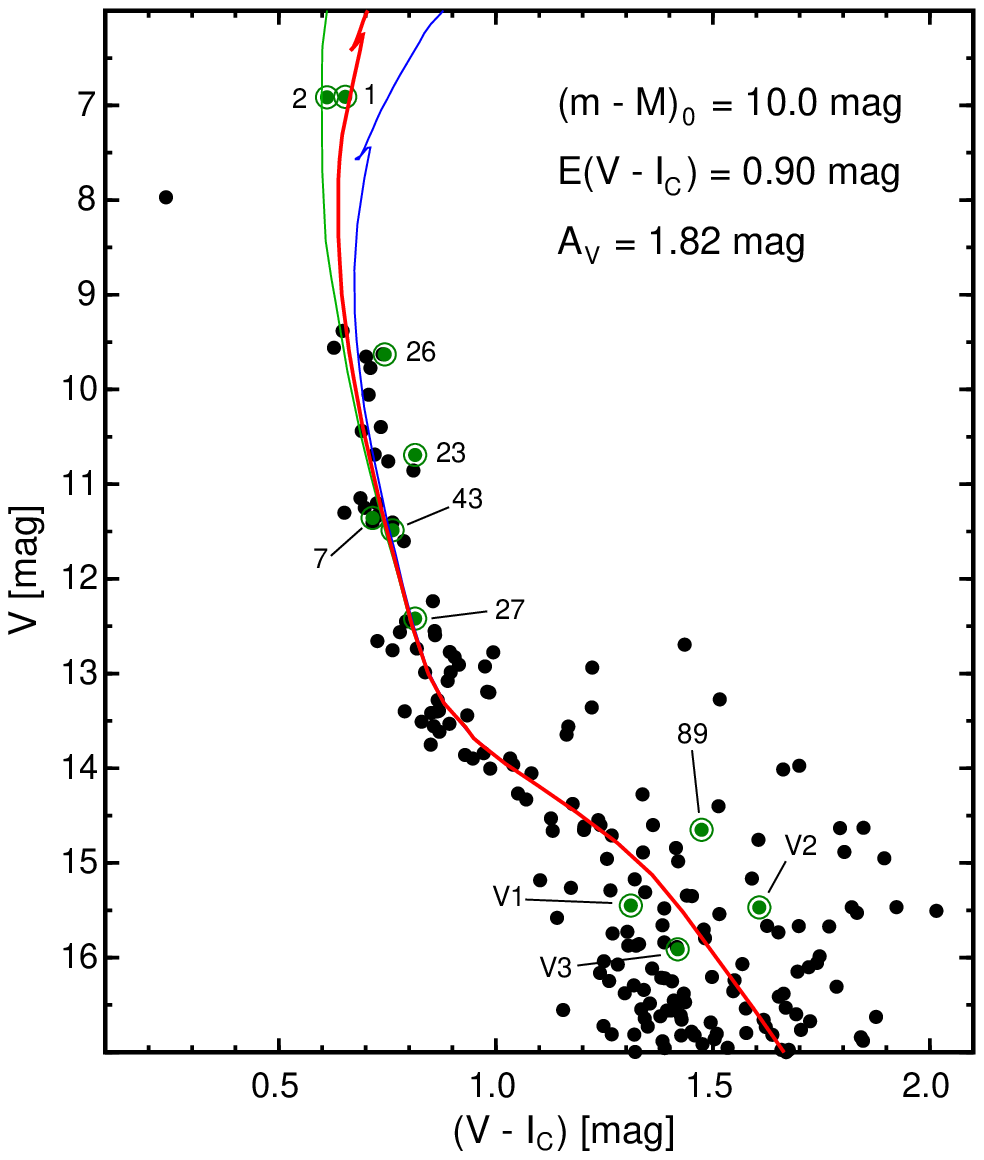}
\FigCap{The $V$ {\it vs.}~$(V-I_{\rm C})$ color-magnitude diagram for NGC\,1502. The eleven variable stars are shown as encircled dots and
labeled. The three isochrones were taken from the models of Bertelli \etal (1994) for $Z =$ 0.019. 
They correspond to $\log(\mbox{age/yr})=$ 6.6, 7.0 and 7.3, i.e.~4, 10 and 20 Myr.}
\label{cmd}
\end{figure}

\begin{figure}[!ht]
\includegraphics[width=10cm]{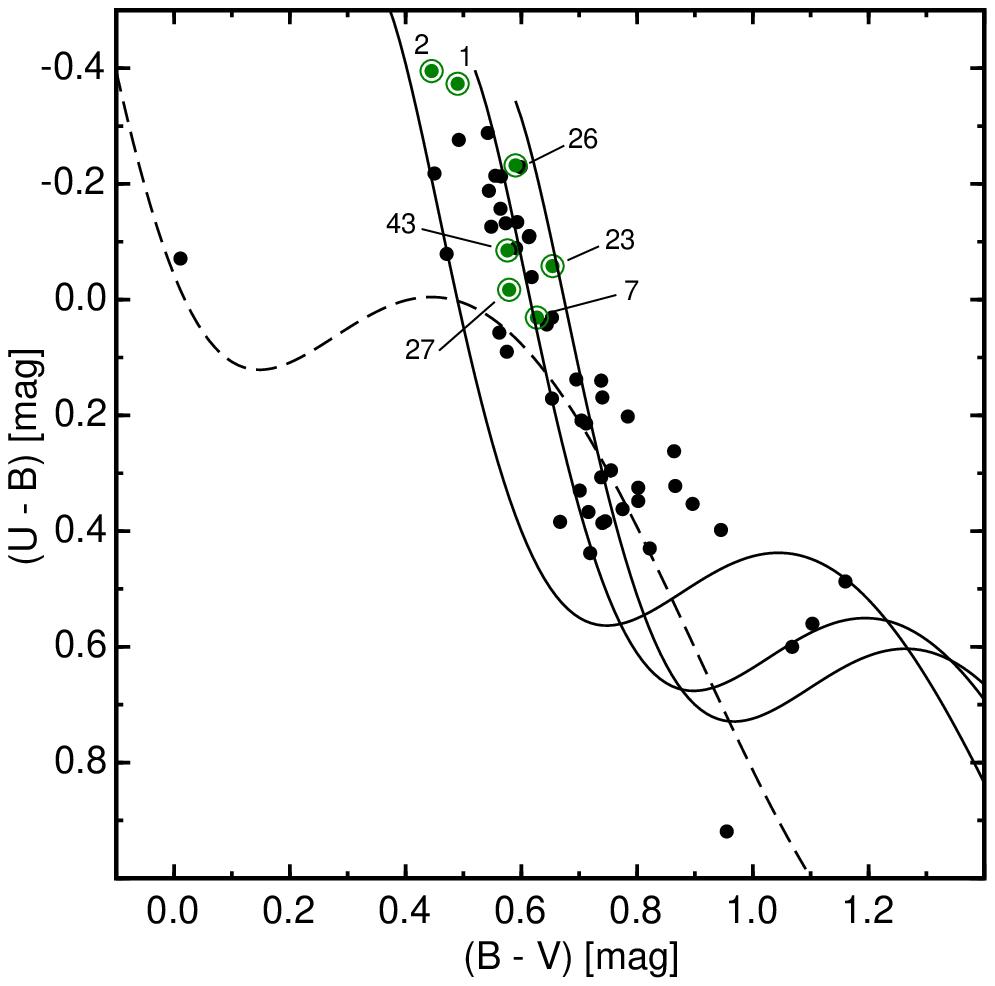}
\FigCap{The $(U-B)$ {\it vs.}~$(B-V)$ color-color diagram for the brightest stars in the observed field of NGC\,1502. The dashed line shows the 
intrinsic color-color relation for main-sequence stars as given by Caldwell \etal (1993).  The same relation
for reddened stars with $E(B-V)=$ 0.60, 0.75 and 0.82~mag is plotted with continuous lines. 
It was assumed that $E(U-B)/E(B-V)=$ 0.724 $+$ 0.021$\times E(B-V)$ (Turner 1989).}
\label{ccd}
\end{figure}

As pointed out in Section 1, the $UBV$ photometry of NGC\,1502 was provided by many authors, but except for Hohle \etal (2009), 
no $I_{\rm C}$ photometry has 
been made so far. In order to obtain standard $(V-I_{\rm C})$ colors, we carried out $VI_{\rm C}$ observations 
of NGC\,1502 together with the Landolt (1992)
field PG\,0918$+$029 on a single clear night of April 20, 2005.  From these data, we derived the $(V-I_{\rm C})$ colors of 29 stars in the central part
of the field.  These stars were subsequently used as secondary standards for the transformation of the instrumental $(V-I_{\rm C})$ colors
obtained with the new camera. The transformation equation was the following:
$$(V-I_{\rm C}) = 0.745(17) \times (v-i_{\rm c}) + 0.739(06),\quad\sigma = 0.029~{\rm mag},$$
where $\sigma$ denotes standard deviation of the fit while numbers in parentheses, the {\it rms} errors of the coefficients with leading zeroes 
omitted. In the transformation equations we use lowercase letters for the instrumental and uppercase for the standard magnitudes.
The $V$ magnitudes, as well as $(U-B)$ and $(B-V)$ colors, were obtained by means of a transformation to the standard system 
using the $UBV$ observations of Hoag \etal (1961), both photoelectric and photographic. The following transformation equations were derived:
$$V = v -0.134(18) \times (v-i_{\rm c}) + 9.628(07),\quad\sigma = 0.043~{\rm mag},$$
$$(U-B) = 1.049(48) \times (u-b) - 0.229(26),\quad\sigma = 0.122~{\rm mag},$$
$$(B-V) = 1.350(30) \times (b-v) + 0.599(08),\quad\sigma = 0.048~{\rm mag}.$$
On average, about 50 stars were used to derive each transformation. The coefficients for color terms considerably different from 1 for 
transformations to $(B-V)$ and $(V-I_{\rm C})$ and different from 0 for transformation to $V$ 
in the above equations are the consequence of the fact that our instrumental $v$ band is shifted towards shorter wavelengths 
with respect to the standard $V$. 
In total, $V$ magnitudes and $(V-I_{\rm C})$ colors were derived for 191 stars in the field, $(B-V)$ colors for 119 stars and $(U-B)$
colors, for 55 stars.  The resulting magnitudes and colors were used to plot the stars from NGC\,1502 in the color-magnitude and color-color 
diagrams (Figs.~\ref{cmd} and \ref{ccd}). The photometry is given in Table 4. In addition, astrometric transformation of mean stellar positions 
derived from images was made in order to obtain equatorial coordinates of the observed stars. As a reference catalogue, we used the USNO\,B1.0 
catalogue (Monet \etal 2003).
A standard deviation of the transformation amounts to about 0.1$^{\prime\prime}$. The positions of stars are also given in Table 4.  
However, we provide here only the entries for variable stars. The full version of Table 4 is available in electronic 
form from the {\it Acta Astronomica Archive}.  

\MakeTable{cccrcccc}{10cm}{$UBVI_{\rm C}$ and H$\alpha$ photometry and coordinates of stars in the field of NGC\,1502}
{\hline\noalign{\vskip1pt}
Star & R.\,Asc.~(2000.0) & Decl.~(2000.0) & $V$ & $(V-I_{\rm C})$ & $(B-V)$ & $(U-B)$ & $\alpha$ \\
\noalign{\vskip1pt}\hline\noalign{\vskip1pt}
 1 & 4$^{\rm h}$07$^{\rm m}$51.38$^{\rm s}$&+62$^{\rm o}$19$^{\prime}$48.4$^{\prime\prime}$& 6.910&$+$0.653&$+$0.490&$-$0.373&2.113\\
 2 & 4$^{\rm h}$07$^{\rm m}$49.27$^{\rm s}$&+62$^{\rm o}$19$^{\prime}$58.7$^{\prime\prime}$& 6.915&$+$0.611&$+$0.445&$-$0.395&2.136\\
26 & 4$^{\rm h}$07$^{\rm m}$44.08$^{\rm s}$&+62$^{\rm o}$18$^{\prime}$04.2$^{\prime\prime}$& 9.630&$+$0.744&$+$0.590&$-$0.232&2.091\\
23 & 4$^{\rm h}$07$^{\rm m}$51.75$^{\rm s}$&+62$^{\rm o}$16$^{\prime}$29.7$^{\prime\prime}$&10.692&$+$0.814&$+$0.654&$-$0.058&2.056\\
43 & 4$^{\rm h}$07$^{\rm m}$49.75$^{\rm s}$&+62$^{\rm o}$22$^{\prime}$25.2$^{\prime\prime}$&11.357&$+$0.716&$+$0.576&$-$0.085&2.193\\
 7 & 4$^{\rm h}$07$^{\rm m}$54.88$^{\rm s}$&+62$^{\rm o}$20$^{\prime}$17.7$^{\prime\prime}$&11.488&$+$0.762&$+$0.627&$+$0.031&2.232\\
27 & 4$^{\rm h}$07$^{\rm m}$40.81$^{\rm s}$&+62$^{\rm o}$15$^{\prime}$40.7$^{\prime\prime}$&12.420&$+$0.814&$+$0.579&$-$0.017&---\\
89 & 4$^{\rm h}$08$^{\rm m}$08.98$^{\rm s}$&+62$^{\rm o}$15$^{\prime}$36.0$^{\prime\prime}$&14.651&$+$1.474&$+$1.453&---&---\\
V1 & 4$^{\rm h}$07$^{\rm m}$23.10$^{\rm s}$&+62$^{\rm o}$24$^{\prime}$14.3$^{\prime\prime}$&15.450&$+$1.311&$+$1.176&---&---\\
V2 & 4$^{\rm h}$07$^{\rm m}$59.92$^{\rm s}$&+62$^{\rm o}$17$^{\prime}$17.6$^{\prime\prime}$&15.470&$+$1.607&$+$1.617&---&1.898\\
V3 & 4$^{\rm h}$07$^{\rm m}$50.82$^{\rm s}$&+62$^{\rm o}$19$^{\prime}$08.2$^{\prime\prime}$&15.912&$+$1.419&---&---&---\\
\noalign{\vskip1pt}
\hline
}

As can be seen from Fig.~\ref{cmd}, the cluster main sequence is well defined, although some scatter ($\approx$0.2 mag)  in $(V-I_{\rm C})$ due to
differential extinction can be seen.  In order to estimate the age of the cluster, we adopted the distance of 1~kpc, \textit{i.e.}, $(m-M)_0=$ 10.0~mag 
and the mean color excess $E(B-V)$ of 0.70~mag, translating into $E(V-I_{\rm C}) \approx$ 0.90~mag and a total absorption in $V$, 
$A_V =$ 1.82~mag ($R_V =$ 2.7 was assumed, see Section 2). All three isochrones we show in Fig.~\ref{cmd} fit the cluster main sequence
equally well and it is the position of the two brightest stars that allows us to constrain the age as the turnoff point of the cluster is poorly
populated.  The comparison allows us to estimate the age of the cluster for 10\,$\pm$\,5~Myr, in a full agreement with previous determinations.

\section{H$\alpha$ Photometry}
\begin{figure}[!ht]
\includegraphics[width=12cm]{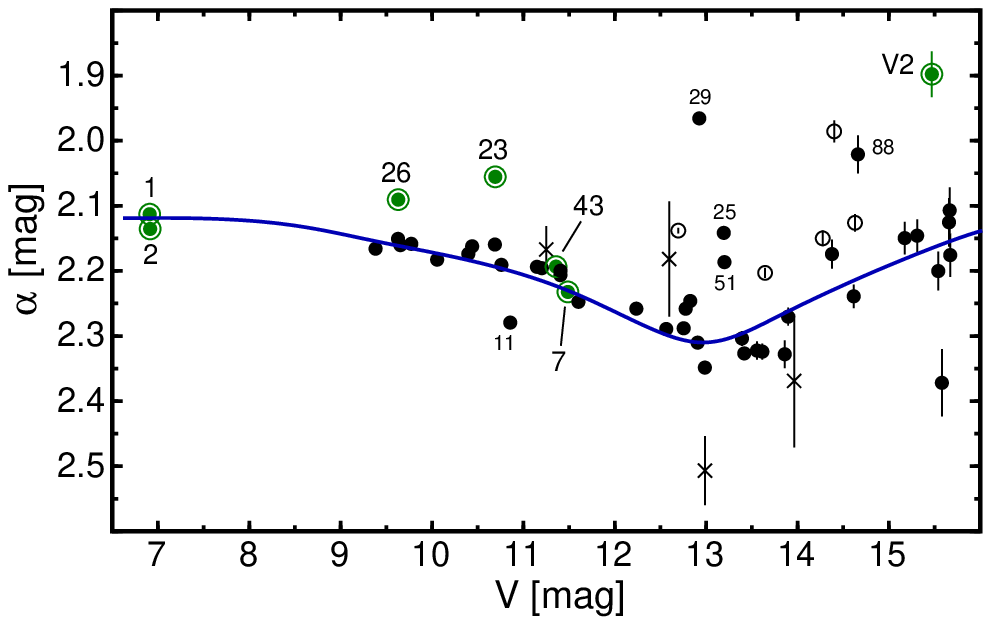}
\FigCap{The $\alpha$ index plotted against $V$ magnitude for the brightest stars in the field covered by the observations made with the old CCD camera.
The four crosses denote relatively faint stars located in the vicinity of the two brightest stars in the field, NGC\,1502-1 and 2. Open circles are 
stars that could be identified as non-members from the color-magnitude diagram (Fig.~\ref{cmd}). The solid line shows calculated location of 
luminosity class V stars that are cluster members; see text for explanation.}
\label{alpha}
\end{figure}

As for most clusters surveyed within our program, we made H$\alpha$ photometry of NGC\,1502 using two filters centered at H$\alpha$, narrow
and wide. The resulting reddening-free $\alpha$ index can be used to find stars with emission at H$\alpha$, including Be stars.  The H$\alpha$ 
observations for NGC\,1502 were carried out with the old camera and therefore the $\alpha$ index could be derived only for the 
brightest stars in the central 
part of the cluster. The indices are shown in Fig.~\ref{alpha} and given in the last column of Table 4.  

Ko{\l}aczkowski \etal (2004) showed how $\alpha$ changes for stars of different spectral types and luminosity classes.  We adopted these 
calculations, but expressed the calculated values of $\alpha$ as a function of absolute magnitude using the absolute magnitude vs.~spectral
type calibration of Schmidt-Kaler (1982). Then, adopting the apparent distance modulus of NGC\,1502, $(m-M)=$ 11.8~mag, we plotted the
resulting $\alpha$ vs.~$V$ relation in Fig.~\ref{alpha} (continuous line).  The relation has a maximum (note the reversed ordinate scale) for
stars with early A spectral type because these stars have the strongest H$\alpha$ lines.  The points follow the relation because
the majority of bright stars belongs to the cluster and have no emission at H$\alpha$.  Stars located well below the relation, like NGC\,1502-11,
are non-members; for those located above it, there are two possibilities: either they are non-members or have
emission at H$\alpha$. The two possibilities are not mutually exclusive, \textit{i.e.}, we can have a non-member with H$\alpha$ in emission. 
However, all stars having $\alpha <$~2.0~mag could be regarded as stars with H$\alpha$ emission. We have three such stars, including 
the variable star V2. Some non-members can be identified from the color-magnitude diagram (Fig.~\ref{cmd}) where they are 
located clearly off the cluster main sequence. Four such stars were plotted in Fig.~\ref{alpha} as open circles. However, there are still several
stars that might have weak emission in H$\alpha$. These are: NGC\,1502-26, the $\beta$~Cep star (see Section 4.1), NGC\,1502-29, 25, 51 and 88.

\section{Conclusions}
The open cluster NGC\,1502 contains only a single low-amplitude $\beta$~Cep star (NGC\,1502-26). This appears to be in a contrast to NGC\,6910
(Ko{\l}aczkowski \etal 2004, Pigulski \etal 2007) which
is not very rich in stars either but harbors as many as seven $\beta$~Cep stars.  Whether the difference can be interpreted in terms of
a difference in metallicity is not known. Unfortunately, the only metallicity determination that can be related to NGC\,1502 is that from 
Str\"omgren photometry by Eggen (1985) for HD\,25056 in Cam~OB1. He reported [Fe/H] $=$ 0.16~dex for this star.  The value of [Fe/H] is rather uncertain. 
In addition, the star might not be related to NGC\,1502 at all. Definitely, a spectroscopic determination of abundances of stars in the cluster 
are highly desirable.

SZ Cam remains one of the most interesting stars of the cluster being an excellent case of a massive hierarchical system that 
are found frequently in the cores of open clusters.  Thorough studies of such systems may shed light on the role of binaries in massive star
formation in general. SZ~Cam was also succesfully used to derive the distance to the cluster (Gorda \etal 2007) and we may hope that further observations
will improve this determination.  The relatively bright EA-type system (NGC\,1502-7) we discovered might be also used for this purpose.

\Acknow{This research has made use of the WEBDA database, operated at the Institute for Astronomy of the University of Vienna.
This work was supported by two grants: N\,N203\,302635 from Polish MNiSzW and Chilean Proyecto FONDECYT Nr 3085010. 
We thank Prof.~M.\,Jerzykiewicz for his comments made upon reading the manuscript.  
We also thank J.\,Molenda-\.Zakowicz, 
G.\,Kopacki and Z.\,Ko{\l}aczkowski for making some observations of the cluster.}

\end{document}